\documentclass[twocolumn]{autart}    

\usepackage{graphicx}          
\usepackage{amsmath,amsfonts,amssymb}
\usepackage{graphicx}
\usepackage{booktabs}
\usepackage{placeins}
\usepackage{algorithm}

\usepackage{algorithmic}
\usepackage{textcomp}
\usepackage{url}

\newtheorem{theorem}{Theorem}
\newtheorem{definition}{Definition}

\newtheorem{assumption}{Assumption}
\newtheorem{lemma}{Lemma}

\newcommand{\E}{\mathbb{E}}
\DeclareMathOperator{\Var}{Var}
\DeclareMathOperator{\tr}{tr}
\newcommand{\R}{\mathbb{R}}

\usepackage{etoolbox}
\providecommand{\qedsymbol}{$\square$}
\AtEndEnvironment{pf}{\unskip\hfill\qedsymbol\par}

\begin{document}

\begin{frontmatter}

\title{Predictive Compensation in Finite--Horizon LQ Games under Gauss--Markov Deviations\thanksref{footnoteinfo}} 

\thanks[footnoteinfo]{This paper was not presented at any IFAC 
meeting. Corresponding author Navid Mojahed. E-mail nmojahed@ucdavis.edu.}

\author[Paestum]{Navid Mojahed}\ead{nmojahed@ucdavis.edu},    
\author[Paestum]{Mahdis Rabbani}\ead{mrabbani@ucdavis.edu},               
\author[Paestum]{Shima Nazari}\ead{snazari@ucdavis.edu}  

\address[Paestum]{Department of Mechanical and Aerospace Engineering, University of California, Davis, CA 95616, USA.}  


\begin{abstract}                          
This paper develops a predictive compensation framework for finite--horizon, discrete--time linear quadratic dynamic games subject to Gauss–-Markov execution deviations from feedback Nash strategies. One player’s control is corrupted by temporally correlated stochastic perturbations modeled as a first--order autoregressive (AR(1)) process, while the opposing player has causal access to past deviations and employs a predictive feedforward strategy that anticipates their future effect. We derive closed--form recursions for mean and covariance propagation under the resulting perturbed closed loop, establish boundedness and sensitivity properties of the equilibrium trajectory, and characterize the reduction in expected cost achieved by optimal predictive compensation. Numerical experiments corroborate the theoretical results and demonstrate performance gains relative to nominal Nash feedback across a range of disturbance persistence levels.
\end{abstract}

\end{frontmatter}

\section{Introduction}
Linear quadratic dynamic games provide a canonical framework for analyzing multi-agent decision-making with coupled dynamics and cost functions. A central concept is the Nash equilibrium, where no player benefits from unilateral deviation~\cite{basar1999dynamic,nash1950equilibrium}. Classical treatments establish the existence and characterization of feedback Nash strategies and coupled Riccati recursions for finite-horizon settings~\cite{engwerda2005lq,LAA_Jungers_2013}, while recent work has revisited scalable or data-driven computation of (approximate) feedback Nash equilibria in discrete time~\cite{nortmann2024,nortmann2022nash}. Beyond model-based schemes, learning-based methods, such as Q-learning tailored to LQ games, have been shown to recover feedback Nash strategies without explicit models; see, for example, the finite-horizon, nonzero-sum difference-game formulation in~\cite{9377313}, which we also use as our numerical benchmark.

Dynamic games serve as foundational models for strategic interactions in multi-agent systems, spanning autonomous driving and motion planning~\cite{chiu2021encoding} to coordination and learning in networked decision systems~\cite{marden2012revisiting}. In such applications, equilibrium feedback policies provide a principled mechanism for anticipating other agents’ actions and maintaining mutual consistency among strategies.

However, when these theoretical policies are deployed on physical systems, implementation imperfections, such as actuator lag, saturation, quantization, and communication delay, can distort the intended control inputs. As a result, the realized behavior may deviate systematically from the nominal Nash trajectory. Existing treatments of uncertainty in dynamic games, including $\varepsilon$-equilibria, robust or risk-sensitive formulations, and soft-constrained formulations~\cite{TANAKA1991413,Jimenez01072006,AMATO2002507_guaranteeing,guerrero2021openloop,Kebriaei2018discrete,van-den-broek2023,wang2022risk}, have primarily focused on model or observation uncertainty and often idealize execution errors as white and memoryless. Yet, in practice, these disturbances are typically temporally correlated, reflecting actuator dynamics or communication holdover, and neglecting this correlation can significantly degrade closed-loop performance and robustness margins, even when the nominal equilibrium ensures internal stability.

Our prior work examined optimal compensation for LQ games under deterministic execution errors~\cite{rabbani2025optimal}, establishing a baseline for quantifying and mitigating equilibrium deviations arising from imperfect actuation. Building on that framework, the present study extends the analysis to stochastic and temporally correlated disturbances modeled as first-order Gauss--Markov (AR(1)) processes. The AR(1) model, a standard representation of colored noise in stochastic control and estimation~\cite{LeMaitre2021GaussMarkov,brown1997introduction,bryson1965linear,shmaliy2022state}, captures the persistence and memory effects inherent in real actuation channels.

Specifically, we analyze a discrete-time, finite-horizon, two--player, nonzero--sum LQ game in which Player 2’s applied input departs from its Nash feedback law by an AR(1) deviation. We derive closed-form recursions characterizing the evolution of second-order moments as the deviation propagates through the time-varying closed loop, revealing analytically that both state covariance and cost inflation scale quadratically with the steady-state per-channel standard deviation, $\mathcal{O}(\sigma_0^2)$, and increase monotonically with the persistence factor~$\rho$. Finally, leveraging the one-step-ahead predictor inherent to the AR(1) structure, we propose a causal predictive compensation strategy for Player 1 that proactively mitigates the expected cost. This compensation strategy is conceptually related to preview and feedforward control~\cite{tomizuka1987zero}, but developed here for game-theoretic LQ settings.

\textbf{Contributions.} 
(i) A moment-based sensitivity framework for discrete-time finite-horizon LQ Nash games subject to AR(1) execution deviations, including explicit coupled recursions for the state covariance and state–deviation cross-covariance.
(ii) Norm- and trace-based bounds that certify $\mathcal{O}(\sigma_0^2)$ scaling of the state covariance and establish monotone growth of cost sensitivity with the persistence factor~$\rho$.
(iii) A causal predictive compensation law for Player~1 that exploits past deviation measurements through the AR(1) structure and, under mild assumptions, strictly reduces the expected cost relative to a no-compensation baseline under the same AR(1) deviations, validated through Monte~Carlo simulations.

\textbf{Paper Outline.} Section~\ref{sec:nash-formulation} formalizes the finite-horizon LQ game and the feedback Nash equilibrium. Section~\ref{sec:structured-exec} introduces the AR(1) execution-deviation model and develops the deviation-propagation recursions and cost-sensitivity bounds. Section~\ref{sec:predictive-design} presents the predictive compensator design and its analysis. Section~\ref{sec:numerical-example} reports numerical results, and Section~\ref{sec: conclusion} concludes with future directions.

\section{Problem Formulation \& Background}
\label{sec:nash-formulation}
Consider a discrete-time, finite-horizon, two-player nonzero-sum linear quadratic (LQ) dynamic game. This section presents the nominal system dynamics, the cost functions of the players, and the resulting feedback Nash equilibrium under perfect execution.

\noindent \textbf{Notation.}
For a matrix $A \in \mathbb{R}^{n\times m}$, the spectral norm is denoted as 
$\|A\|_2 := \sigma_{\max}(A)$, where $\sigma_{\max}(A)$ is the largest singular value of $A$, and the Frobenius norm is represented by 
$\|A\|_F := \sqrt{\mathrm{tr}(A^\top A)}$, where $\mathrm{tr}(\cdot)$ denotes the trace. 
For symmetric $A,B$, we write $A \succ 0$ ($\succeq 0$) if $A$ is positive (semi)definite, and $A \succ B$ ($\succeq B$) if $A-B$ is positive (semi)definite. The spectral radius of a matrix is denoted by $\varrho(\cdot)$. 
An asterisk $^*$ denotes that the variable belongs to the nominal Nash equilibrium; for instance, $x_k^*$, $u_{i,k}^*$, and $J_i^*$, represent the nominal Nash equilibrium state, control, and cost for Player~$i$, respectively. 
The subscript $_{-i}$ refers to all players except Player~$i$. 
For a random variable or stochastic process, $\E[\cdot]$ and $\Var[\cdot]$ denote its expectation and variance, respectively.

\noindent \textbf{Game Model.}
We study the discrete-time LQ dynamic game
\begin{equation}
    x_{k+1} = A x_k + B_1 u_{1,k} + B_2 u_{2,k}, 
    \quad k = 0, \ldots, N-1,
    \label{eq:state-dynamics}
\end{equation}
where $x_k \in \mathbb{R}^{n}$ is the joint states of the game, $u_{i,k} \in \mathbb{R}^{m_i}$ is the control of Player~$i\in\{1,2\}$, and the initial joint states $x_0$ is given. The system matrices satisfy $A\in\mathbb{R}^{n\times n}$, and $B_i\in\mathbb{R}^{n\times m_i}$, with $n$ number of joint states and $m_i$ number of the control inputs of Player~$i$.

Each player $i$ minimizes the quadratic cost
\begin{equation}
    J_i = \sum_{k=0}^{N-1} \big( x_k^\top Q_i x_k + u_{i,k}^\top R_i u_{i,k} \big) 
          + x_N^\top Q_{i,N} x_N,
    \label{eq:cost-function}
\end{equation}
where $Q_i, \,Q_{i,N}\succeq 0$, and $R_i\succ 0$ are stage and terminal weights.

\begin{assumption}\label{assump:weights}
For each $i\in\{1,2\}$, the weighting matrices satisfy $Q_i, \,Q_{i,N}\succeq 0$, and $R_i\succ 0$. 
The positive definiteness of $R_i$ ensures well-posedness of the Nash feedback law.
\end{assumption}

\begin{definition}
A pair of contol inputs $\{(u^*_{1,k}, \, u^*_{2,k})\}_{k=0}^{N-1}$ with
\begin{equation}
    u^*_{i,k} = -K^*_{i,k}\,x_k,\quad i\in\{1,2\},
    \label{eq:feedback-nash}
\end{equation}
is a feedback Nash equilibrium if, for each Player~\(i\),
\begin{equation*}
\begin{aligned}
J_i\bigl(x_0,\{u^*_{i,k}\}_{k=0}^{N-1},&\{u^*_{-i,k}\}_{k=0}^{N-1}\bigr)\;\\ &\le\;
J_i\bigl(x_0,\{u_{i,k}\}_{k=0}^{N-1},\{u^*_{-i,k}\}_{k=0}^{N-1}\bigr),
\end{aligned}
\end{equation*}
for all admissible control sequences \(\{u_{i,k}\}_{k=0}^{N-1}\).
\end{definition}

Under Nash equilibrium gains \(\{K^*_{1,k}, \, K^*_{2,k}\}_{k=0}^{N-1}\), the closed-loop system evolves as
\begin{equation}
x_{k+1}^* = A_{\mathrm{cl},k}\,x_k^*,\quad
A_{\mathrm{cl},k} := A - B_1 K^*_{1,k} - B_2 K^*_{2,k}.
\label{eq:Acl-def}
\end{equation}

\begin{assumption}
\label{assump:nash-existence}
For the finite-horizon LQ game in \eqref{eq:state-dynamics}–\eqref{eq:cost-function}, a unique feedback Nash equilibrium exists if:
\begin{enumerate}
    \item Assumption~\ref{assump:weights} holds,
    \item The pair $(A,\bar B)$, with $\bar B := [B_1\; B_2]$ is controllable.
    \item The coupled Riccati difference equations admit a unique solution for all $k$, which is guaranteed if
    \[
    M_k :=
    \begin{bmatrix}
    I_n & B_1 R_1^{-1} B_1^\top & B_2 R_2^{-1} B_2^\top \\
    P_{1,k} & -I_n & 0 \\
    P_{2,k} & 0 & -I_n
    \end{bmatrix},
    \]
    is invertible, where $I_n$ denotes the $n\times n$ identity and $P_{i,k}$ are the Riccati matrices of Player~$i$.
\end{enumerate}
These conditions are standard in finite-horizon LQ game theory~\cite{basar1999dynamic,van-den-broek2023,LAA_Jungers_2013} and are typically satisfied for fully actuated, well-posed systems.
\end{assumption}

Given Assumptions~\ref{assump:weights}--\ref{assump:nash-existence}, the equilibrium gain matrices \(\{K^*_{1,k}, \, K^*_{2,k}\}_{k=0}^{N-1}\) are obtained via coupled Riccati difference equations. Each player’s value function is quadratic, $V_{i,k}(x)=x^\top P_{i,k} x$, with terminal condition $P_{i,N}=Q_{i,N}$.
At each stage $k$, the equilibrium feedback laws \eqref{eq:feedback-nash} follow from jointly solving the coupled Riccati recursion (see~\cite{basar1999dynamic,engwerda2005lq} for explicit forms). 
These coupled recursions guarantee existence and uniqueness of the feedback Nash gains under Assumptions~\ref{assump:weights}--\ref{assump:nash-existence}.

In the next section, we extend this nominal formulation by incorporating structured execution deviations in Player~2’s control input, modeled as a temporally correlated stochastic process.

\section{Structured Execution Deviations \& Moment--Based Sensitivity Analysis}
\label{sec:structured-exec}
In practical implementation, the Nash feedback strategy may not be executed exactly due to actuator imperfections, communication latency, embedded-controller quantization, or other hardware nonlinearities. To model such deviations, we assume that Player~2’s implemented input differs from its nominal feedback Nash law,
\begin{equation} \label{eq:delta_u_2_in_u*}
    u_{2,k} = u_{2,k}^* + \Delta u_{2,k}, \qquad
    u_{2,k}^* = -K^*_{2,k} x_k,
\end{equation}
where $\Delta u_{2,k}\in\mathbb{R}^{m_2}$ represents execution-level error.

Under this model, the perturbed closed-loop dynamics evolve as
\begin{equation}
    x_{k+1} = A_{\mathrm{cl},k} x_k + B_2 \Delta u_{2,k},
    \label{eq:perturbed-dyn}
\end{equation}
where $A_{\mathrm{cl},k}\in\mathbb{R}^{n\times n}$ is the time-varying closed-loop matrix defined in~\eqref{eq:Acl-def}. 
Here, $B_2\,\Delta u_{2,k}$ acts as a colored stochastic excitation that propagates through the closed-loop system. 

\begin{assumption}\label{assump:ar1}
The deviation sequence $\{\Delta u_{2,k}\}_{k\ge 0}$ follows a first-order Gauss--Markov (AR(1)) recursion
\begin{equation}\label{eq:ar1}
\Delta u_{2,k} \;=\; \rho\,\Delta u_{2,k-1} \;+\; \sigma_w\, w_k, \qquad k\ge 1,
\end{equation}
with $\Delta u_{2,0}=0$, scalar persistence factor $0\leq\rho<1$, and i.i.d.\ innovations $w_k\sim\mathcal N(0,I_{m_2})$, independent of $x_0$. 
We adopt the variance-preserving parameterization
\begin{equation}\label{eq:sigmaw}
\sigma_w \;=\; \sqrt{1-\rho^2}\,\sigma_0,
\end{equation}
so that the marginal variance converges to $\sigma_0^2 I_{m_2}$, i.e. $\Var(\Delta u_{2,k})\to \sigma_0^2 I_{m_2}$, for some prescribed steady-state per-channel standard deviation~$\sigma_0>0$.
Player~1 has causal access to $\Delta u_{2,k-1}$ (or equivalently, can recover it via $\Delta u_{2,k}=u_{2,k}+K_{2,k}^*x_k$ when $u_{2,k}$ is observed).
For analytical tractability, $\rho$ is taken to be common across channels \cite{LeMaitre2021GaussMarkov}.
\end{assumption}




We examine how Player~2’s structured execution deviation \eqref{eq:ar1} propagates through the time-varying closed-loop system~\eqref{eq:perturbed-dyn} and affects the second-order moments that determine Player~1’s cost. Let \(x_k^*\) denote the nominal state trajectory under the feedback Nash, and define
\begin{equation}
    \Delta x_k := x_k - x_k^*.
\end{equation}
Subtracting the nominal dynamics \eqref{eq:Acl-def} from the perturbed ones~\eqref{eq:perturbed-dyn} yields
\begin{equation}\label{eq:error-dynamics}
    \Delta x_{k+1} = A_{\mathrm{cl},k}\,\Delta x_k + B_2\,\Delta u_{2,k},
    \qquad \Delta x_0 = 0,
\end{equation}
with expectations, $\E[\cdot]$, taken with respect to the innovation sequence $\{w_k\}$ in \eqref{eq:ar1}.


\begin{lemma}\label{lem:AR1-cov}
Under Assumption~\ref{assump:ar1}, define
\[
\Phi_k := \mathbb{E}[\Delta u_{2,k}\Delta u_{2,k}^\top],\qquad
\Phi_{k,\ell} := \mathbb{E}[\Delta u_{2,k}\Delta u_{2,\ell}^\top].
\]
Then, for all $k\ge \ell\ge0$
\begin{align}
&\Phi_{k,\ell} = \rho^{\,k-\ell}\,\Phi_\ell, \label{eq:AR1-cross-cov}\\
&\Phi_k = \rho^2 \Phi_{k-1} + \sigma_w^2 I_{m_2},\qquad \Phi_0 = 0. \label{eq:AR1-marginal}
\end{align}
In particular, the marginal covariance admits a closed form
\begin{equation}\label{eq:AR1-closed-form}
\Phi_k \;=\; \sigma_0^2\,(1-\rho^{2k})\,I_{m_2}\ \ \preceq\ \ \sigma_0^2 I_{m_2},
\end{equation}
and satisfies $\Phi_{k,\ell}^\top=\Phi_{\ell,k}$. Hence $\Phi_k \to \sigma_0^2 I_{m_2}$ as $k\to\infty$.
\end{lemma}
\begin{pf}
Fix $\ell\ge 0$ and proceed by induction on $k\ge \ell$. The base case $k=\ell$ is immediate:
$\Phi_{\ell,\ell}=\E[\Delta u_{2,\ell}\Delta u_{2,\ell}^\top]=\Phi_\ell=\rho^{\ell-\ell}\Phi_\ell$. 

Assume $\Phi_{k,\ell}=\rho^{k-\ell}\Phi_\ell$ holds for some $k\ge \ell$. Using the AR(1) update  \eqref{eq:ar1}, we obtain
\[
\begin{aligned}
\Phi_{k+1,\ell}
&= \E[\Delta u_{2,k+1}\Delta u_{2,\ell}^\top] \\
&= \rho\,\E[\Delta u_{2,k}\Delta u_{2,\ell}^\top]
  + \sigma_w\,\E[w_{k+1}\Delta u_{2,\ell}^\top].
\end{aligned}
\]
The random variable $w_{k+1}$ is independent of $\Delta u_{2,\ell}$ and has a zero mean; thus, the second term is zero. This yields
$\Phi_{k+1,\ell}= \rho\,\Phi_{k,\ell}
= \rho\,\rho^{k-\ell}\Phi_\ell
= \rho^{k+1-\ell}\Phi_\ell$,
which completes the induction and proves \eqref{eq:AR1-cross-cov}.

For the marginal covariance, from $\Delta u_{2,k}=\rho\Delta u_{2,k-1}+\sigma_w w_k$,
\begin{align*}
\Phi_k
&= \E\!\big[(\rho\,\Delta u_{2,k-1}+\sigma_w w_k)(\rho\,\Delta u_{2,k-1}+\sigma_w w_k)^\top\big] \nonumber\\
&= \rho^2\,\E[\Delta u_{2,k-1}\Delta u_{2,k-1}^\top]
 + \sigma_w^2\,\E[w_k w_k^\top]  \nonumber\\
&\quad + \rho\sigma_w\,\E[\Delta u_{2,k-1}w_k^\top]
 + \rho\sigma_w\,\E[w_k\Delta u_{2,k-1}^\top].
\end{align*}
The cross terms vanish (independence and zero mean), and $\E[w_k w_k^\top]=I_{m_2}$. Thus
\begin{equation*}
\Phi_k=\rho^2\Phi_{k-1}+\sigma_w^2 I_{m_2},\qquad \Phi_0=0,
\end{equation*}
which is \eqref{eq:AR1-marginal}.

Unfolding the scalar linear recursion \eqref{eq:AR1-marginal} and using \eqref{eq:sigmaw} yields
\begin{equation*}
    \Phi_k=\sum_{j=0}^{k-1}\rho^{2j}\sigma_w^2 I_{m_2}
= \sigma_0^2(1-\rho^{2k})\,I_{m_2}
\preceq \sigma_0^2 I_{m_2},
\end{equation*}
establishing \eqref{eq:AR1-closed-form} and the uniform bound. Symmetry follows by definition:
\begin{equation*}
    \Phi_{k,\ell}^\top=\mathbb{E}[(\Delta u_{2,k}\Delta u_{2,\ell}^\top)^\top]=\mathbb{E}[\Delta u_{2,\ell}\Delta u_{2,k}^\top]=\Phi_{\ell,k}.
\end{equation*}
Finally, since $0\leq\rho<1$, we have $\rho^{2k}\to 0$; thus, $\Phi_k\to \sigma_0^2 I_{m_2}$. This completes the proof.\end{pf}

With the deviation second-order statistics characterized by Lemma~\ref{lem:AR1-cov}, we now propagate these statistics through the deviation dynamics \eqref{eq:error-dynamics}. Introducing the state–deviation moments $(\Sigma_k,C_k)$, we obtain closed-form linear recursions that quantify how colored input disturbance $B_2\Delta u_{2,k}$ shapes the state covariance under the time-varying closed loop. The next theorem formalizes these recursions and establishes uniform boundedness.

\begin{theorem}\label{thm:mean-covariance}
Building on Lemma~\ref{lem:AR1-cov}, let $\Delta u_{2,k}\in\mathbb{R}^{m_2}$ follow Assumption~\ref{assump:ar1} with $\sigma_w=\sqrt{1-\rho^2}\,\sigma_0$, and let the deviation-driven dynamics be
\[
\Delta x_{k+1}=A_{\mathrm{cl},k}\,\Delta x_k+B_2\Delta u_{2,k},\qquad \Delta x_0=0,
\]
Assume $\{A_{\mathrm{cl},k}\}$ is uniformly Schur stable, i.e., there exist constants $c\ge1$ and $\beta<1$ such that $\|A_{\mathrm{cl},k}^j\|_2 \le c\,\beta^j$ for all $j,k$. Define the second-order moments
\begin{equation}\label{eq:Sigma-Phi-C-def}
\begin{aligned}
\Sigma_k &:= \E[\Delta x_k \Delta x_k^\top], \\
\Phi_k   &:= \E[\Delta u_{2,k} \Delta u_{2,k}^\top], \\
C_k      &:= \E[\Delta x_k \Delta u_{2,k}^\top].
\end{aligned}
\end{equation}

Then $\E[\Delta x_k]=0$ for all $k\ge0$, and the moments satisfy
\begin{align}
\Sigma_{k+1} &= A_{\mathrm{cl},k}\,\Sigma_k\,A_{\mathrm{cl},k}^\top
  + B_2 \Phi_k B_2^\top + A_{\mathrm{cl},k} C_k B_2^\top \label{eq:sigma-rec}\\
  &+ B_2 C_k^\top A_{\mathrm{cl},k}^\top, 
    \nonumber  \\
C_{k+1} &= \rho\,A_{\mathrm{cl},k}\,C_k + \rho\,B_2 \Phi_k,\qquad C_0=0,
  \label{eq:C-rec}
\end{align}
with $\Phi_k$ from Lemma~\ref{lem:AR1-cov} \eqref{eq:AR1-marginal}.

All matrices $\Sigma_k$, $\Phi_k$, and $C_k$ are uniformly bounded in $k$, satisfying $\|\Sigma_k\|_2=\mathcal{O}(\sigma_0^2)$.
\end{theorem}
\begin{pf}
We analyze the evolution of the second-order deviation moments induced by the AR(1) execution perturbation. Recall \eqref{eq:error-dynamics} under Assumption~\ref{assump:ar1}, and i.i.d.\ zero-mean Gaussian ${w_k}$, independent across time and from the initial state. We establish: (i) $\E[\Delta x_k]=0$, (ii) the moment recursions \eqref{eq:sigma-rec}–\eqref{eq:C-rec}, and (iii) uniform $\mathcal{O}(\sigma_0^2)$ bounds.

\noindent \textbf{(i) Zero-mean property.}
The AR(1) recursion and $\E[w_k]=0$ imply $\E[\Delta u_{2,k}]=0$ for all $k$ (by induction). Taking expectations in the deviation dynamics gives
\[
\mathbb{E}[\Delta x_{k+1}]
= A_{\mathrm{cl},k}\,\mathbb{E}[\Delta x_k]+B_2\,\mathbb{E}[\Delta u_{2,k}]=A_{\mathrm{cl},k}\,\mathbb{E}[\Delta x_k].
\]
Since $\E[\Delta x_0]=0$, induction yields $\E[\Delta x_k]=0$ for all $k$.



\noindent \textbf{(ii) Covariance recursions.}
Consider the second-order moments defined in \eqref{eq:Sigma-Phi-C-def}.
Expanding the deviation dynamics,
\begin{equation*}
\begin{aligned}
\Delta &x_{k+1}\Delta x_{k+1}^\top\\
&= \big(A_{\mathrm{cl},k}\Delta x_k+B_2\Delta u_{2,k}\big)
   \big(A_{\mathrm{cl},k}\Delta x_k+B_2\Delta u_{2,k}\big)^\top \\
&= A_{\mathrm{cl},k}\Delta x_k\Delta x_k^\top A_{\mathrm{cl},k}^\top
 + B_2\Delta u_{2,k}\Delta u_{2,k}^\top B_2^\top \\
&\quad + A_{\mathrm{cl},k}\Delta x_k\Delta u_{2,k}^\top B_2^\top
 + B_2\Delta u_{2,k}\Delta x_k^\top A_{\mathrm{cl},k}^\top.
\end{aligned}
\end{equation*}
Taking expectations yields \eqref{eq:sigma-rec}.

Using the AR(1) update for the next deviation,
\[
\Delta u_{2,k+1}=\rho\,\Delta u_{2,k}+\sigma_w w_{k+1},
\]
we obtain
\[
\begin{aligned}
C_{k+1}
&:=\E[\Delta x_{k+1}\Delta u_{2,k+1}^\top] \\
&= \E\big[(A_{\mathrm{cl},k}\Delta x_k+B_2\Delta u_{2,k})
         (\rho\,\Delta u_{2,k}+\sigma_w w_{k+1})^\top\big] \\
&= \rho\,A_{\mathrm{cl},k}\,\E[\Delta x_k\Delta u_{2,k}^\top]
  + \rho\,B_2\,\E[\Delta u_{2,k}\Delta u_{2,k}^\top] \\
&\quad + \sigma_w A_{\mathrm{cl},k}\,\E[\Delta x_k w_{k+1}^\top]
  + \sigma_w B_2\,\E[\Delta u_{2,k} w_{k+1}^\top].
\end{aligned}
\]
The last two terms vanish since $w_{k+1}$ is zero-mean and independent of $(\Delta x_k,\Delta u_{2,k})$. Thus
\[
C_{k+1}=\rho\,A_{\mathrm{cl},k}C_k+\rho\,B_2\Phi_k,\qquad C_0=0,
\]
which is \eqref{eq:C-rec}. Eventually, Lemma~\ref{lem:AR1-cov} provides $\Phi_k$.

\noindent \textbf{(iii) Uniform $\mathcal{O}(\sigma_0^2)$ bounds.}
Unroll the linear deviation dynamics:
\begin{align*}
\Delta x_k
&= \sum_{j=0}^{k-1}
   \Big(A_{\mathrm{cl},k-1}\cdots A_{\mathrm{cl},j+1}\Big)B_2\,\Delta u_{2,j}\\
&=: \sum_{j=0}^{k-1} G_{k,j}\,\Delta u_{2,j},
\end{align*}
with $G_{k,j} \in \mathbb{R}^{n\times m_2}$. By convention, $\prod_{t=a}^{b} A_{\mathrm{cl},t}=I$ when $a>b$ (so $G_{k,k-1}=B_2$).
Thus
\[
C_k
= \E[\Delta x_k\Delta u_{2,k}^\top]
= \sum_{j=0}^{k-1} G_{k,j}\,\E[\Delta u_{2,j}\Delta u_{2,k}^\top].
\]
Using Lemma~\ref{lem:AR1-cov}, for $j\le k$,
\begin{equation*}
    \E[\Delta u_{2,j}\Delta u_{2,k}^\top]=\rho^{\,k-j}\Phi_j, \qquad ||\Phi_j||_2\leq\sigma_0^2.
\end{equation*}
Taking spectral norms and using submultiplicativity,
\begin{equation*}
\|C_k\|_2
\le \sum_{j=0}^{k-1} \|G_{k,j}\|_2\,\rho^{\,k-j}\,\|\Phi_j\|_2.
\end{equation*}
By the uniform Schur stability assumption, there exist constants $c\ge 1$ and $\beta\in(0,1)$ such that
$\|A_{\mathrm{cl},t_1}\cdots A_{\mathrm{cl},t_r}\|_2 \le c\,\beta^{r}$ for all admissible products; hence
\[
\|G_{k,j}\|_2\le c\,\beta^{\,k-1-j}\,\|B_2\|_2.
\]
As \(\|\Phi_j\|_2\le\sigma_0^2\) for all \(j\),
\begin{equation}
    \label{eq:C-bound}
\begin{aligned}
\|C_k\|_2
&\le c\,\|B_2\|_2\,\sigma_0^2
\sum_{j=0}^{k-1} \beta^{\,k-1-j}\rho^{\,k-j} \\
&= c\,\|B_2\|_2\,\sigma_0^2\,\rho
   \sum_{m=0}^{k-1}(\beta\rho)^m
\;\le\; \frac{c\,\|B_2\|_2\,\rho}{1-\beta\rho}\,\sigma_0^2,
\end{aligned}
\end{equation}
since $0\leq\rho<1$ and $\beta<1$ imply $\beta\rho<1$.

From \eqref{eq:sigma-rec} and submultiplicativity,
\[
\begin{aligned}
\|\Sigma_{k+1}\|_2
&\le \beta^2\,\|\Sigma_k\|_2
 + \|B_2\|_2^2\,\|\Phi_k\|_2 \\
&\quad + \|A_{\mathrm{cl},k}\|_2\,\|C_k\|_2\,\|B_2\|_2\\
& \quad+ \|B_2\|_2\,\|C_k\|_2\,\|A_{\mathrm{cl},k}\|_2 \\
&\le \beta^2\,\|\Sigma_k\|_2
 + \|B_2\|_2^2\,\sigma_0^2
 + 2\beta\,\|B_2\|_2\,\|C_k\|_2.
\end{aligned}
\]
Let
\begin{equation}\label{eq:C1-def}
C_1\;:=\;\frac{2\,c\,\beta\,\rho}{1-\beta\rho}\,.
\end{equation}
Then, using the bound \eqref{eq:C-bound}, we have
\begin{align*}
2\beta\,\|B_2\|_2\,\|C_k\|_2
\;&\le\; 2\beta\,\|B_2\|_2\,
\frac{c\,\|B_2\|_2\,\rho}{1-\beta\rho}\,\sigma_0^2\\
&\;=\; C_1\,\|B_2\|_2^2\,\sigma_0^2,
\end{align*}
Therefore,
\[
\|\Sigma_{k+1}\|_2
\;\le\; \beta^2\,\|\Sigma_k\|_2
\;+\; \big(1+C_1\big)\,\|B_2\|_2^2\,\sigma_0^2.
\]
Unrolling from $\Sigma_0=0$ yields
\[
\|\Sigma_k\|_2
\;\le\; \big(1+C_1\big)\,\|B_2\|_2^2\,\sigma_0^2
\sum_{j=0}^{k-1}\beta^{2j}
\;\le\; C_2\,\sigma_0^2,
\]
where
\begin{equation}\label{eq:c2-def}
C_2\;:=\;\frac{\big(1+C_1\big)\,\|B_2\|_2^2}{1-\beta^2}\,.
\end{equation}
Hence \(\sup_k \|\Sigma_k\|_2 \le C_2\,\sigma_0^2\), so \(\sup_k \|\Sigma_k\|_2 = \mathcal{O}(\sigma_0^2)\).
Together with \(\sup_k\|\Phi_k\|_2\le\sigma_0^2\) and  \(\sup_k\|C_k\|_2=\mathcal{O}(\sigma_0^2)\), completes the proof. 
\end{pf}

\begin{rem}
The recursion \eqref{eq:sigma-rec} shows that the state covariance $\Sigma_k$ depends not only on the marginal deviation covariance $\Phi_k$ but also on the cross-covariance $C_k$, which encodes temporal correlation.
As the persistence factor~$\rho$ increases, past deviations decay more slowly, amplifying $C_k$ (cf.~\eqref{eq:C-rec}) and, through the mixed terms in~\eqref{eq:sigma-rec}, enlarging $\Sigma_k$. 
Under uniform closed-loop stability ($\|A_{\mathrm{cl},k}\|_2\le\beta<1$ up to a constant $c$), $\Sigma_k$ remains bounded with $\mathcal{O}(\sigma_0^2)$ scaling.
Equations~\eqref{eq:C1-def} and~\eqref{eq:c2-def} demonstrate that the bound constant $C_2$ grows monotonically with $\rho$, illustrating how correlation persistence degrades closed-loop robustness even without instability.
\end{rem}

The derived moment bounds reveal a fundamental limitation of nominal Nash feedback strategies: their inability to mitigate temporally correlated execution errors. Even under uniformly Schur-stable closed-loop dynamics, state covariances and cost increments increase monotonically with $\rho$. 
This degradation arises because the nominal Nash policy, optimized for memoryless disturbances, ignores the temporal structure encoded in $C_k$.

Exploiting the Gauss--Markov (AR(1)) structure enables predictive compensation, wherein past deviations are used to anticipate and counteract expected future errors and cost increases. We focus on a scalar AR(1) model with scalar persistence factor~$\rho$, which provides a parsimonious yet effective model in independent control channels. This simplification allows closed-form cost propagation analysis while capturing the dominant effects of structured execution imperfections in practical systems.

\section{Predictive Compensator Design}
\label{sec:predictive-design}
This section designs a causal predictive compensator for Player~1 that exploits the Gauss--Markov structure of Player~2's execution deviation to not only attenuate its predictable component, but also potentially take advantage of it. The information available to Player~1 at time $k$ is
\[
\mathcal I_k := \sigma\big(x_0,\{x_t\}_{t=0}^k,\{u_{1,t}\}_{t=0}^{k-1},\{\Delta u_{2,t}\}_{t=0}^{k-1}\big),
\]
and admissible policies satisfy $u_{1,k}=\mu_k(\mathcal I_k)$.

\begin{definition}
\label{def:predictive-feedback}
Under Assumption~\ref{assump:ar1} with $0\leq\rho<1$ and $\sigma_w$ as in~\eqref{eq:sigmaw}, the one-step conditional predictor of Player~2’s deviation is
\begin{equation*}
\E\!\big[\Delta u_{2,k}\,\big|\,\Delta u_{2,k-1}\big] \;=\; \rho\,\Delta u_{2,k-1}.
\end{equation*}
Leveraging this prediction, Player~1 applies the $\mathcal I_k$--measurable preview feedback policies of the form
\begin{equation}\label{eq:predictive-feedback}
u_{1,k} \;=\; -K^*_{1,k}\,x_k \;-\; L_k\,\rho\,\Delta u_{2,k-1},
\end{equation}
where \(K^*_{1,k}\) is the nominal feedback Nash gain and $L_k\in\R^{m_1\times m_2}$ is a compensation gain to be chosen to minimize the expected cost increment \(\E[\Delta J_1]\). This causal law exploits the autoregressive structure to proactively adjust Player~1’s input and improve robustness to temporally correlated execution errors.
\end{definition}

\begin{theorem}
\label{thm:predictive_feedback_general}
Consider the game \eqref{eq:state-dynamics}–\eqref{eq:cost-function} under Assumptions~\ref{assump:weights} and \ref{assump:ar1} with $0\leq\rho<1$ and $\sigma_w$ as in \eqref{eq:sigmaw}.
Let $\Delta J_1:=J_1-J_1^*$ denote Player~1’s cost increment. Under a frozen-covariance approximation, i.e., treat the second moments
\[
\Phi_{k-1}:=\mathbb{E}[\Delta u_{2,k-1}\Delta u_{2,k-1}^\top],\qquad
C_k:=\mathbb{E}[\Delta x_k\,\Delta u_{2,k}^\top]
\]
as fixed (their nominal values with $L_k\equiv0$), the stagewise minimizer of the quadratic approximation of $\mathbb{E}[\Delta J_1]$ over $L_k$ is
\begin{equation}
\label{eq:Lk-star}
L_k^* \;=\; -\,\frac{1}{\rho^{2}}\,K^*_{1,k}\,C_k\,\Phi_{k-1}^{-1}.
\end{equation}
where $\Phi_{k-1}^{-1}$ is interpreted on the support of $\Phi_{k-1}$ (or taken as the Moore–Penrose pseudoinverse~\cite{benisrael2003generalized} when necessary). Substituting \eqref{eq:Lk-star} into \eqref{eq:predictive-feedback} yields the equivalent form
\begin{equation}
    \label{eq:preview-law-compact}
    u_{1,k} \;=\; -\,K^*_{1,k}\!\left(x_k - \frac{1}{\rho}\,C_k\,\Phi_{k-1}^{-1}\,\Delta u_{2,k-1}\right).
\end{equation}
If $\rho>0$ and $C_k\neq0$ on a set of nonzero measure, then $\E[\Delta J_1]$ is strictly reduced relative to $L_k\equiv0$. For $k=0$, $z_0=0$ and $L_0$ is immaterial; set $L_0=0$ without loss of generality.
\end{theorem}
\begin{pf}
Let $z_k := \rho\,\Delta u_{2,k-1}$ ($z_0=0$), so $\Delta u_{2,k}=z_k+\sigma_w w_k$ with $\E[z_k]=0$, $\E[w_k]=0$. Under \eqref{eq:predictive-feedback}, $\Delta u_{1,k}=-K^*_{1,k}\Delta x_k - L_k z_k$.
Expanding the LQ cost yields 
\begin{align*}
\Delta J_1
&= \sum_{k=0}^{N-1}\!\Big(\Delta x_k^\top Q_1 \Delta x_k
+ \Delta u_{1,k}^\top R_1 \Delta u_{1,k}\Big)\\
& \quad+ \Delta x_N^\top Q_1 \Delta x_N.
\end{align*}
Taking expectations and under the frozen-covariance approximation, we can collect the terms depending on $L_k$.
Define
\[
\Psi_k := \E[z_k z_k^\top] = \rho^2\,\Phi_{k-1}\succeq 0,\quad
\Gamma_k := \E[\Delta x_k z_k^\top] = C_k.
\]
Then
\begin{equation*}
\E[\Delta J_1] \;=\; \sum_{k=0}^{N-1} J_k(L_k) \;+\; \text{const},
\end{equation*}
where the stage-wise part, depending on $L_k$ is quadratic
\begin{equation}\label{eq:Jk-Lk}
J_k(L_k)
= \tr\!\big(L_k^\top R_1 L_k\,\Psi_k\big)
  + 2\,\tr\!\big(L_k^\top R_1 K^*_{1,k}\,\Gamma_k\big),
\end{equation}
and \textbf{\text{const}} collects $\tr(Q_1\Sigma_k)$, $\tr(K_{1,k}^{*\top}R_1K^*_{1,k}\Sigma_k)$, and the terminal terms independent from $L_k$.

Since $R_1\succ 0$ and $\Psi_k\succeq 0$, $J_k$ is convex in $L_k$, and the first-order optimality gives
\[
2R_1 L_k \Psi_k + 2R_1 K^*_{1,k}\Gamma_k = 0
\;\;\Longrightarrow\;\;
L_k^* = -\,K^*_{1,k}\,\Gamma_k\,\Psi_k^{-1}.
\]
Using $\Gamma_k=C_k$ and $\Psi_k=\rho^2\Phi_{k-1}$ results in \eqref{eq:Lk-star}.

Substituting this into \eqref{eq:predictive-feedback} yields the compact form \eqref{eq:preview-law-compact}.

Furthermore, completing the square stated in \eqref{eq:Jk-Lk} shows $J_k(L_k)-J_k(L_k^*)=\tr((L_k-L_k^*)^\top R_1 (L_k-L_k^*)\,\Psi_k)\ge0$, with strict inequality when $\rho>0$ and $C_k\neq 0$. Therefore $J_k(L_k^*)\le J_k(L_k)$ for any $L_k$, in particular
$J_k(L_k^*)\le J_k(0)$. Summing  over $k$ yields
\begin{equation*}
    \E[\Delta J_1(L^*)] \le \E[\Delta J_1(L)],\quad
\E[\Delta J_1(L^*)] \le \E[\Delta J_1(0)].
\end{equation*}
and completes the proof.\end{pf}
The law \eqref{eq:preview-law-compact} cancels the predictable component of Player~2’s deviation seen through $C_k$, making the residual control perturbation orthogonal (in the $R_1$-weighted sense) to the predictable disturbance. This mirrors classical preview/feedforward compensation in LQ tracking, here adapted to the game-theoretic setting.

\section{Numerical Example}
\label{sec:numerical-example}
The purpose of the numerical study is twofold:
(i) to validate the analytical moment recursions for the deviation dynamics derived from
\eqref{eq:perturbed-dyn}–\eqref{eq:ar1}, and
(ii) to quantify how the steady-state per-channel standard deviation~$\sigma_0$ and persistence factor~$\rho$
affect the deviation energy and the resulting cost under both nominal and predictive laws
\eqref{eq:feedback-nash} and \eqref{eq:predictive-feedback}.

We consider the finite-horizon LQ game defined by \eqref{eq:state-dynamics}–\eqref{eq:cost-function} with horizon $N=9$ and dimensions $n=3$ and $m_1=m_2=3$.
The system and weighting matrices are chosen as follows:
\begin{equation*}
\begin{aligned}
    A &= \begin{bmatrix}
0.7484 & 0.2386 & 0.0703\\
0.2386 & 0.6585 & 0.2795\\
0.0703 & 0.2795 & 0.4471
\end{bmatrix},\\
B_1 &= \begin{bmatrix}
0.3561 & 0.0820 & 0.0905\\
0.0820 & 0.3496 & 0.2702\\
0.0905 & 0.2702 & 0.3838
\end{bmatrix},\\[2pt]
B_2 &= \begin{bmatrix}
0.2748 & 0.0950 & 0.0965\\
0.0950 & 0.3439 & 0.2671\\
0.0965 & 0.2671 & 0.3797
\end{bmatrix},
\end{aligned}
\end{equation*}
\begin{equation*}
\begin{gathered}
Q_1 = Q_2 =
\begin{bmatrix}
50 & 5 & 2\\
5 & 10 & 3\\
2 & 3 & 1
\end{bmatrix},\qquad
Q_{1,N} = Q_{2,N} = Q_1,\\
R_1 = R_2 =
\begin{bmatrix}
15.0 & 2.25 & 0.0\\
2.25 & 7.5  & 0.0\\
0.0  & 0.0  & 3.0
\end{bmatrix},
\end{gathered}
\end{equation*}
which satisfy Assumption~\ref{assump:weights}. 

\begin{figure}[t]
    \centering
    \includegraphics[width=0.95\linewidth]{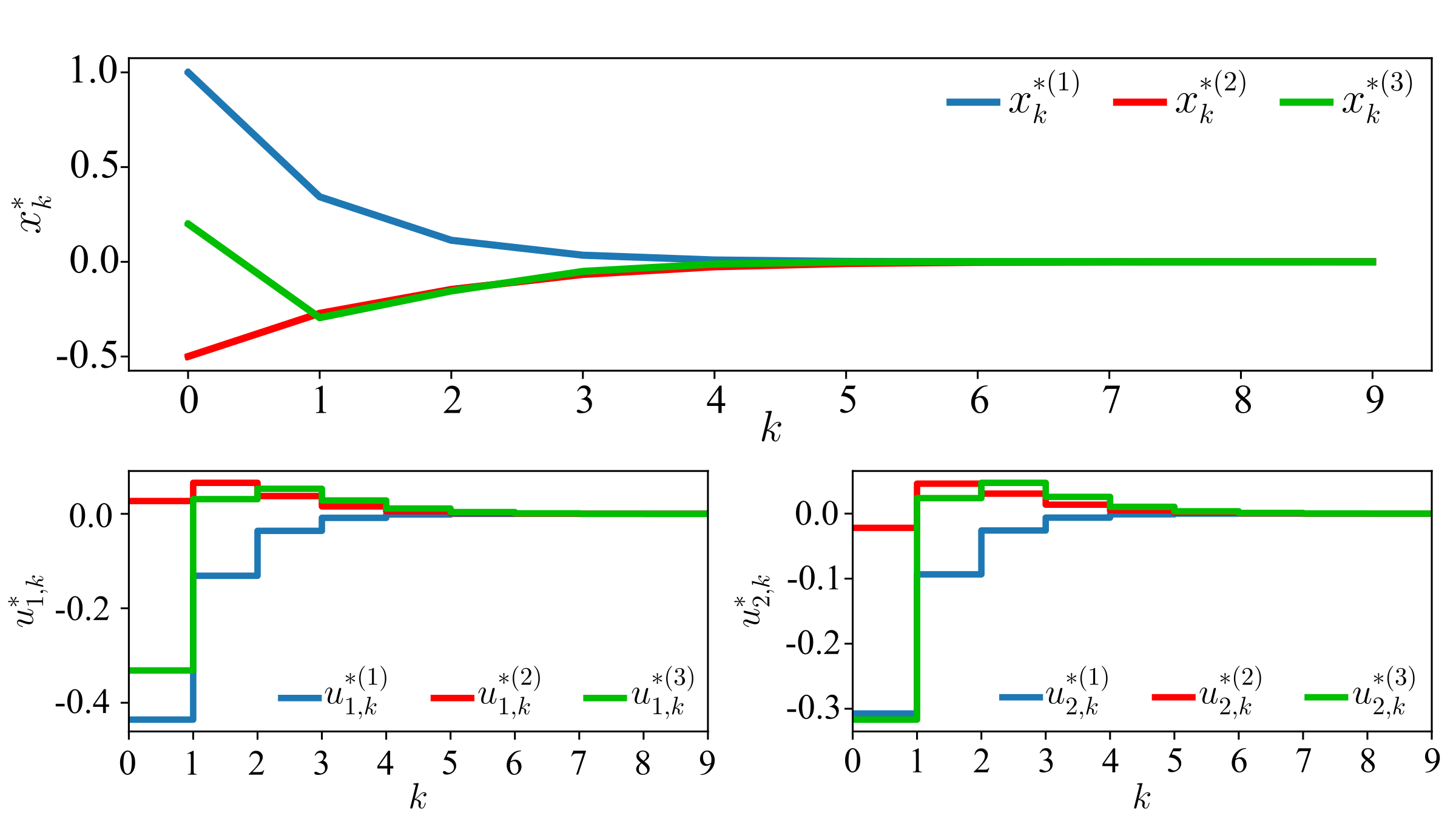}
    \caption{Nominal feedback–Nash trajectories for both players. 
    \textbf{Top}: state components $x_k^{(i)}$, $i=1,2,3$. 
    \textbf{Bottom}: corresponding control inputs $u_{1,k}^{(i)}$ (left) and $u_{2,k}^{(i)}$ (right) under the equilibrium policies ${K_{1,k}^*,K_{2,k}^*}$. The stable closed-loop evolution serves as the reference for all subsequent deviation experiments.}
    \label{fig:nominal-trajs}
\end{figure}

\noindent \textbf{Nominal Feedback Nash.}
Solving the coupled Riccati recursions yields stage-wise feedback Nash gains $\{K^*_{1,k},K^*_{2,k}\}_{k=0}^{N-1}$ and closed-loop matrices $A_{\mathrm{cl},k}$ as in~\eqref{eq:Acl-def}. All stages are Schur-stable with $\max_k \varrho(A_{\mathrm{cl},k})\!\approx\!0.36<1$, ensuring well-posed moment propagation. The resulting nominal trajectories are shown in Fig.~\ref{fig:nominal-trajs} and are used as the baseline throughout the rest of this section.

To model execution imperfections, we perturb Player~2’s Nash law input \eqref{eq:feedback-nash} using the first-order Gauss--Markov (AR(1)) model \eqref{eq:ar1}, with the scaling $\sigma_w=\sqrt{1-\rho^2}\,\sigma_0$ chosen such that $\Var(\Delta u_{2,k})\to\sigma_0^2 I$.
The parameter~$\sigma_0>0$ controls the steady-state per-channel standard deviation, while the persistence factor~$\rho\in[0,1)$ encodes the disturbance memory or temporal correlation.

Consistent with the information structure of Section \ref{sec:structured-exec},
Player~1 has causal access to the previous deviation $\Delta u_{2,k-1}$ and employs the predictive law \eqref{eq:predictive-feedback} with optimal gain $L^*_k$ from \eqref{eq:Lk-star} in Theorem~\ref{thm:predictive_feedback_general}.

The numerical study has two goals:
\begin{enumerate}
    \item Validate Theorem~\ref{thm:mean-covariance} by checking, against Monte Carlo (MC), the second-moment recursions \eqref{eq:sigma-rec}–\eqref{eq:C-rec}, the zero-mean property $\E[\Delta x_k]=0$, boundedness of $\Sigma_k$, and the $\mathcal{O}(\sigma_0^2)$ scaling.
    \item Validate Theorem~\ref{thm:predictive_feedback_general} by comparing the expected cost under (a) nominal feedback Nash ($L_k\equiv 0$) and (b) predictive compensation ($L_k=L_k^*$), across various $\sigma_0$ and $\rho$ values.
\end{enumerate}

\noindent \textbf{Monte Carlo (MC) Simulation.} 
For each pair $(\rho,\sigma_0)$, we run $M=500$ independent MC trials. The sweep grid is
\begin{equation*}
    \begin{aligned}
        \rho\in\{0,0.2,0.4,0.6,0.8,0.9\},\\
        \sigma_0\in\{0.02,0.04,0.06,0.08\}.
    \end{aligned}
\end{equation*}
Each trial simulates the perturbed closed loop for $N$ time steps starting from $x_0=0$, $\Delta u_{2,0}=0$. We record the state deviations $\Delta x_k$, form empirical moments
\[
\widehat{\Sigma}_k=\tfrac{1}{M}\sum_{m=1}^M \Delta x_k^{(m)}\Delta x_k^{(m)\top},\quad
\widehat{\E}[\Delta x_k]=\tfrac{1}{M}\sum_{m=1}^M \Delta x_k^{(m)},
\]
and compute the expected costs $\widehat{\E}[J_1]$ and reductions $\widehat{\E}[\Delta J_1]$. Algorithm~\ref{alg:mc-eval} demonstrates the full procedure of our Monte Carlo simulation.
\begin{algorithm}[t]
\caption{Monte Carlo Evaluation}
\label{alg:mc-eval}
\begin{algorithmic}[1]
\REQUIRE System matrices $(A,B_1,B_2)$, cost weights $(Q_i,R_i,Q_{i,N})$, horizon $N$, $(\rho,\sigma_0)$, trials $M$.
\STATE Compute nominal Nash gains $K^*_{1,k},K^*_{2,k}$ via the coupled Riccati recursion, and calculate $A_{\mathrm{cl},k}$ \eqref{eq:Acl-def}.
\FOR{$m=1$ to $M$}
    \STATE Initialize $\{\tilde{x}_0, x_0,\Delta u_{2,0},\Phi_0,C_0,L^*_0 , \zeta_0\}=0$.
    \FOR{$k=0$ to $N-1$}
        \STATE Compute $\tilde{x}_{k+1}$ \eqref{eq:perturbed-dyn} and $\tilde{u}_{1,k+1}$ \eqref{eq:feedback-nash}.
        \STATE Apply Player~1 law $u_{1,k} = -K^*_{1,k}x_k - L_k\,\zeta_k$ \eqref{eq:predictive-feedback}.
        \STATE Apply Player~2 input $u_{2,k}=u_{2,k}^*+\Delta u_{2,k}$ \eqref{eq:delta_u_2_in_u*}.
        \STATE Propagate $x_{k+1}=A x_k+B_1 u_{1,k}+B_2 u_{2,k}$ \eqref{eq:state-dynamics}.
        \STATE Sample $w_{k+1}\sim\mathcal N(\mathbf{0},I_{m_2})$.
        \STATE Compute $\Delta u_{2,k+1}=\rho\,\Delta u_{2,k}+\sigma_w w_{k+1}$ \eqref{eq:ar1}.
        \STATE Update $\zeta_{k+1} \leftarrow \rho\,\Delta u_{2,k}$.
        \STATE Compute $C_{k+1} $ via~\eqref{eq:C-rec}.
        \STATE Update $L_{k+1}^* \leftarrow -\frac{1}{\rho^2}K^*_{1,k+1}C_{k+1}\Phi_{k}^{-1}$.
        \STATE Compute $\Phi_{k+1} $ via \eqref{eq:AR1-marginal}.
    \ENDFOR
    \STATE Compute $J_1^{(m)}$ and $\tilde{J}^{(m)}_1$from \eqref{eq:cost-function}.
\ENDFOR
\STATE Calculate $\widehat{\E}[J_1]$ from $\{J_1^{(m)}\}_{m=1}^M$.
\STATE Calculate $\widehat{\E}[\Delta J_1]=\widehat{\E}[J_1-\tilde{J}_1]$.
\end{algorithmic}
\end{algorithm}

\begin{figure}[t]
  \centering
  \includegraphics[width=\linewidth]{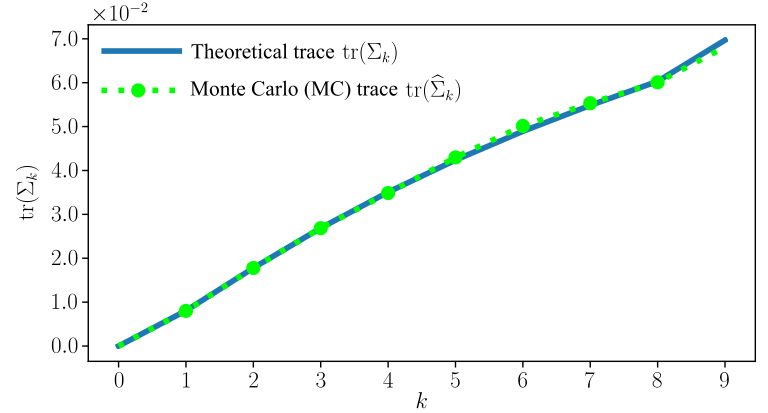}
  \caption{Validation of the second-moment recursion in Theorem~\ref{thm:mean-covariance}. 
  Comparison between theoretical and Monte Carlo (MC) traces $\mathrm{tr}(\Sigma_k)$ and $\mathrm{tr}(\widehat{\Sigma}_k)$ across the horizon.
  The near-perfect overlap confirms the accuracy of the analytical covariance propagation and its boundedness under the Schur-stable closed loop.}
  \label{fig:sigma_trace}
\end{figure}

Figure~\ref{fig:sigma_trace} compares $\mathrm{tr}(\Sigma_k)$ from the recursions \eqref{eq:sigma-rec} in Theorem~\ref{thm:mean-covariance} against its MC estimate $\mathrm{tr}(\widehat{\Sigma}_k)$ over the horizon; the curves coincide almost exactly across all time steps, demonstrating that the derived moment recursions faithfully capture the empirical second-order behavior of the stochastic deviations. 
Moreover, the covariance remains uniformly bounded throughout the horizon, consistent with Schur stability of $A_{\mathrm{cl},k}$ and the boundedness claim of Theorem~\ref{thm:mean-covariance}.

Table~\ref{tab:covariance-scaling} quantifies the growth of $\max_k\mathrm{tr}(\Sigma_k)$ as $\sigma_0$ increases.  
The nearly quadratic scaling, approximately $\{4,9,16\}\!\times$ for $\sigma_0$ doubled, tripled, and quadrupled, empirically confirms the $\mathcal{O}(\sigma_0^2)$ law predicted by Theorem~\ref{thm:mean-covariance}.
\begin{table}[t]
  \centering
  \caption{Empirical verification of the $\mathcal{O}(\sigma_0^2)$ scaling in Theorem~\ref{thm:mean-covariance}.
  The second column reports the absolute $\max_k \mathrm{tr}(\Sigma_k)$ over the horizon; the rightmost column shows the ratio relative to the baseline at $\sigma_0{=}0.15$.}
  \label{tab:covariance-scaling}
  \begin{tabular}{c c c}
    \toprule
    $\sigma_0$ & $\max_k\,\mathrm{tr}(\Sigma_k)$ & Ratio to baseline\\
    \midrule
    0.15 & $9.802\times 10^{-3}$ & Reference\\
    0.30 & $3.921\times 10^{-2}$ & $\approx 4.00\times$\\
    0.45 & $8.822\times 10^{-2}$ & $\approx 9.00\times$\\
    0.60 & $1.568\times 10^{-1}$ & $\approx 16.00\times$\\
    \bottomrule
  \end{tabular}
\end{table}

\begin{figure*}[t]
  \centering
  \includegraphics[width=\linewidth]{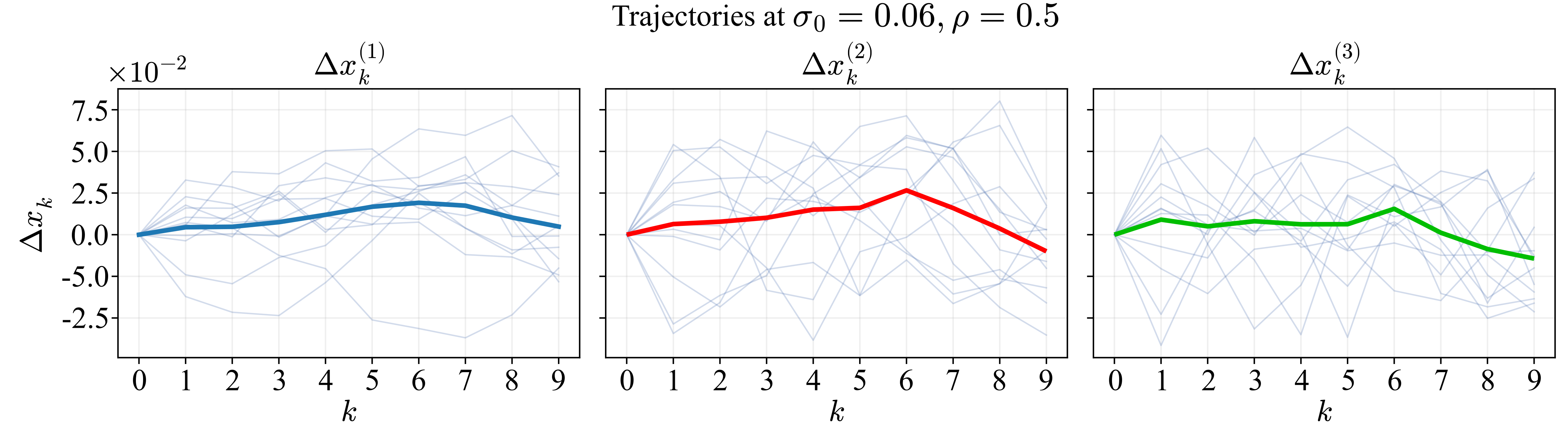}
  \caption{Monte Carlo realizations of the deviation state $\Delta x_k$ at $(\sigma_0,\rho)=(0.06,0.5)$. 
  Faded light gray lines show individual runs; bold curves denote componentwise means.
  Trajectories remain zero-mean and bounded, illustrating the empirical validity of Theorem~\ref{thm:mean-covariance} and the anisotropic spread induced by $(A,B_2)$ and $A_{\mathrm{cl},k}$.}
  \label{fig:deltax-multi}
\end{figure*}

Figure~\ref{fig:deltax-multi} plots multiple Monte Carlo realizations of $\Delta x_k$ for $(\sigma_0,\rho)=(0.06,0.5)$.  
Trajectories fluctuate around zero mean (bold lines), meaning that the empirical means $\widehat{\E}[\Delta x_k]$ are zero, and remain bounded over the horizon.  
These qualitative properties confirm $\E[\Delta x_k]=0$ and finite $\Sigma_k\succeq0$, again in full agreement with Theorem~\ref{thm:mean-covariance}.  
The anisotropic spreads across components arise from the directional amplification induced by the interaction of $(A,B_2)$ and the time-varying $A_{\mathrm{cl},k}$.

We now evaluate the predictive compensation law~\eqref{eq:predictive-feedback} and~\eqref{eq:Lk-star} and its impact on Player~1’s expected cost under temporally correlated deviations.
For each $(\rho,\sigma_0)$, two controllers are compared:
(i) the nominal Nash policy ($L_k\equiv0$) and
(ii) the compensated policy ($L_k=L_k^*$).

\begin{figure*}[t]
    \centering
    \includegraphics[width=\linewidth]{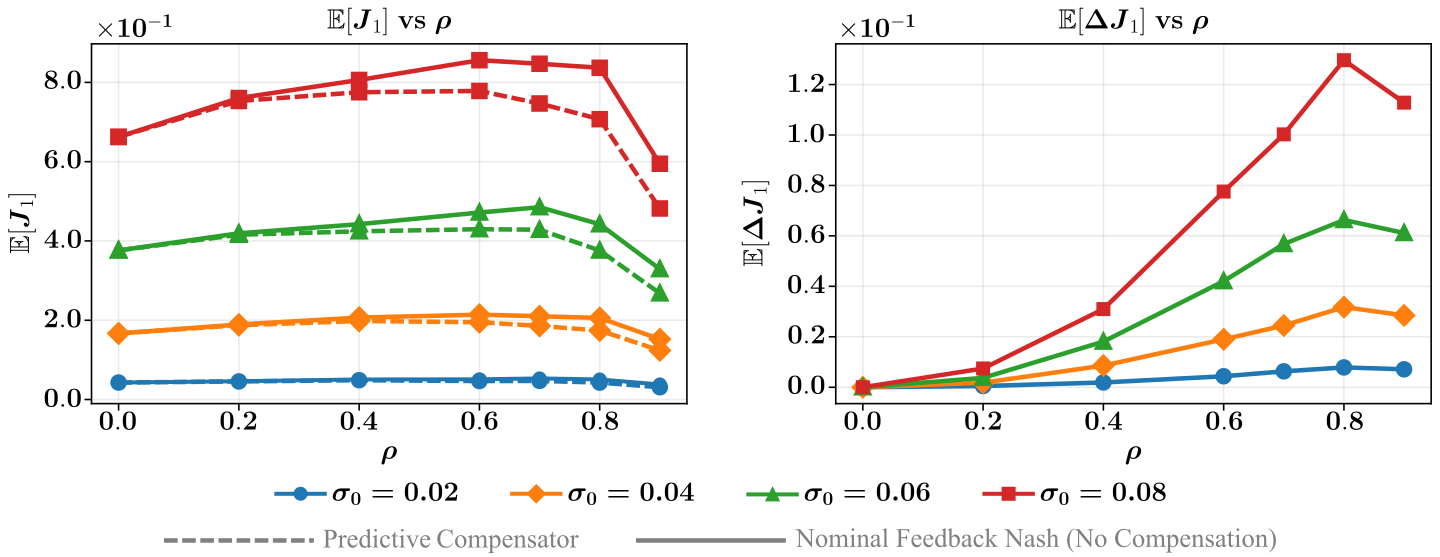}
    \caption{Effect of disturbance persistence and predictive compensation on Player~1’s cost.
    Left: expected cost $\mathbb{E}[J_1]$ versus persistence factor~$\rho$ across various steady-state per-channel standard deviations ~$\sigma_0$.
    Right: expected cost reduction $\mathbb{E}[\Delta J_1]$ due to the predictive compensator.
    Dashed curves (predictive compensator) consistently lie below solid ones (no compensation).}
    \label{fig:cost-and-delta-vs-rho}
\end{figure*}

Figure~\ref{fig:cost-and-delta-vs-rho} summarizes the results.  
The left panel shows that $\E[J_1]$ increases monotonically with both $\rho$ and $\sigma_0$: larger and more persistent execution errors raise the cost, as expected.  
For all $\rho>0$, the predictive compensator yields strictly lower costs than the uncompensated feedback Nash, while at $\rho=0$ (white noise) the two coincide as Theorem~\ref{thm:predictive_feedback_general} predicts.

The right panel plots $\E[\Delta J_1]$ versus $\rho$.  
The reduction grows monotonically with $\rho$ and scales roughly with $\sigma_0^2$, consistent with the quadratic sensitivity of Theorem~\ref{thm:mean-covariance}.  
Across the grid, relative gains are modest ($<3\%$) but systematic, highlighting that exploiting even weak temporal correlation can yield tangible cost savings, whereas purely memoryless noise offers no benefit to prediction.

The mild peak in the cost reduction near large $\rho$ in Fig.~\ref{fig:cost-and-delta-vs-rho} (right) is explained by two competing effects. 
First, stronger temporal correlation amplifies the deviation–state cross-covariance via the recursion \eqref{eq:C-rec}, which pushes $\E[\Delta J_1]$ upward. 
Second, with the variance-preserving scaling \eqref{eq:sigmaw}, the marginal deviation covariance obeys the closed form \eqref{eq:AR1-closed-form}, so the early-stage deviation energy is small when $\rho\to 1$ (for $\rho=1-\varepsilon$ and small $k$, $\Phi_k\approx 2k(1-\rho)\sigma_0^2$). 
Over a finite horizon, these effects yield a peak at large $\rho$ followed by a slight drop as $\rho\to 1$. 
A useful rule of thumb places the peak where the AR(1) warm-up time $\sim 1/(1-\rho)$ matches the horizon, i.e.,
\[
\rho_{\text{peak}}\approx 1-\frac{1}{N},
\]
which $\rho_{\text{peak}}\approx 0.89$ for $N=9$.
Moreover, as $\rho\to 1$ the one-step predictor $\E[\Delta u_{2,k}\mid\Delta u_{2,k-1}]=\rho\,\Delta u_{2,k-1}$ becomes nearly exact, so the predictive compensator cancels most of the deviation, reinforcing the observed downturn.

\section{Conclusion}
\label{sec: conclusion}

This paper presented a sensitivity analysis framework for finite-horizon discrete-time linear–quadratic (LQ) dynamic games subject to structured deviations from the nominal feedback Nash equilibrium. Execution imperfections were modeled through a first-order Gauss--Markov autoregressive (AR(1)) process applied to one player's control input, thereby capturing temporally correlated errors representative of actuator imperfections such as delays and persistent disturbances commonly observed in real systems.

The analytical developments established how such deviations propagate through the closed-loop Nash dynamics and perturb the cost of the opposing player.  
Explicit moment recursions were derived for the state covariance and the state–deviation cross-covariance, leading to trace- and norm-based bounds that certify bounded deviation energy and quantify its $\mathcal{O}(\sigma_0^2)$ scaling.  
Building upon this structure, a causal predictive compensation law was formulated, augmenting the nominal feedback Nash with a feedforward term based on measured deviation history.  
The resulting controller admits a closed-form optimal gain and guarantees a strict reduction in expected cost whenever temporal correlation ($\rho>0$) is present.  
Monte Carlo simulations verified the theoretical moment recursions, confirmed the scaling and boundedness properties, and demonstrated consistent performance improvements under temporally correlated disturbances.

\paragraph*{\textbf{Future Directions.}}
Two extensions appear particularly promising:
\begin{enumerate}
  \item \textbf{Matrix-valued memory.}
  Generalizing the scalar persistence factor to a matrix $\rho\in\mathbb{R}^{m_2\times m_2}$ would allow inter-channel coupling in the deviation dynamics.  
  The resulting recursion $\Phi_{k+1}=\rho\,\Phi_k\rho^\top+\sigma_w^2 I$ remains Lyapunov-like and bounded under spectral stability, enabling higher-fidelity modeling and preserving tractable covariance propagation.

  \item \textbf{Fractional-order deviations.}
  Modeling long-range dependence (long-memory effects) via fractional AR/Gauss--Markov processes~\cite{durand2022fiarma,SCALAS2000376} can capture heavy-tailed or slowly decaying correlation structures.  
  With modern numerical solvers for fractional equations~\cite{mojahed2020finite}, this direction is computationally feasible and may reveal new robustness–performance trade-offs in dynamic games with persistent uncertainty.
\end{enumerate}

Beyond full-information settings, integrating predictive compensation with learning-based opponent modeling constitutes a natural next step.  
Embedding the proposed structure into adaptive or data-driven Nash seeking frameworks, such as PACE and its nonlinear extension N-PACE~\cite{soltanian2025paceframeworklearningcontrol,soltanian2025peerawarecostestimationnonlinear}, could maintain near-Nash performance when opponents’ policies or cost parameters are only partially known.


\bibliographystyle{plain}        
\bibliography{autosam}           

@book{basar1999dynamic,
  author = {T. Basar and G. J. Olsder},
  title = {Dynamic Noncooperative Game Theory},
  publisher = {SIAM},
  year = {1999}
}

@article{nash1950equilibrium,
  author = {J. F. Nash},
  title = {Equilibrium points in n-person games},
  journal = {Proceedings of the National Academy of Sciences},
  volume = {36},
  number = {1},
  pages = {48--49},
  year = {1950}
}

@article{nortmann2024,
  author = {B. Nortmann and A. Monti and M. Sassano and T. Mylvaganam},
  title = {Nash Equilibria for Linear Quadratic Discrete-Time Dynamic Games via Iterative and Data-Driven Algorithms},
  journal = {IEEE Transactions on Automatic Control},
  volume = {69},
  number = {10},
  pages = {6561--6575},
  year = {2024}
}

@article{LAA_Jungers_2013,
  author    = {Marc Jungers},
  title     = {Feedback Strategies for Discrete-Time Linear-Quadratic Two-Player Descriptor Games},
  journal   = {Linear Algebra and its Applications},
  volume    = {439},
  number    = {10},
  pages     = {3111--3134},
  year      = {2013},
  publisher = {Elsevier},
  doi       = {10.1016/j.laa.2013.07.015},
  url       = {https://doi.org/10.1016/j.laa.2013.07.015}
}

@ARTICLE{Kebriaei2018discrete,
  author={Kebriaei, Hamed and Iannelli, Luigi},
  journal={IEEE Transactions on Automatic Control}, 
  title={Discrete-Time Robust Hierarchical Linear-Quadratic Dynamic Games}, 
  year={2018},
  volume={63},
  number={3},
  pages={902-909},
  doi={10.1109/TAC.2017.2719158}}

@article{AMATO2002507_guaranteeing,
author = {Amato, F. and Mattei, M. and Pironti, A.},
title = {Guaranteeing cost strategies for linear quadratic differential games under uncertain dynamics},
journal = {Automatica},
volume = {38},
number = {3},
pages = {507-515},
year = {2002},
issn = {0005-1098},
doi = {https://doi.org/10.1016/S0005-1098(01)00224-2},
url = {https://www.sciencedirect.com/science/article/pii/S0005109801002242}
}

@article{Jimenez01072006,
author = {Manuel Jimenez and Alex Poznyak},
title = {$\varepsilon$-Equilibrium in LQ Differential Games with Bounded Uncertain Disturbances: Robustness of Standard Strategies and New Strategies with Adaptation},
journal = {International Journal of Control},
volume = {79},
number = {7},
pages = {786--797},
year = {2006},
publisher = {Taylor \& Francis},
doi = {10.1080/00207170600690624},
URL = {https://doi.org/10.1080/00207170600690624},
eprint = {https://doi.org/10.1080/00207170600690624}
}

@article{TANAKA1991413,
title = {On $\varepsilon$-equilibrium point in a noncooperative n-person game},
journal = {Journal of Mathematical Analysis and Applications},
volume = {160},
number = {2},
pages = {413-423},
year = {1991},
issn = {0022-247X},
doi = {https://doi.org/10.1016/0022-247X(91)90314-P},
url = {https://www.sciencedirect.com/science/article/pii/0022247X9190314P},
author = {Kensuke Tanaka and Kazunori Yokoyama}
}

@inproceedings{chiu2021encoding,
  title={Encoding defensive driving as a dynamic Nash game},
  author={Chiu, Chih-Yuan and Fridovich-Keil, David and Tomlin, Claire J},
  booktitle={2021 IEEE International Conference on Robotics and Automation (ICRA)},
  pages={10749--10756},
  year={2021},
  organization={IEEE}
}

@book{engwerda2005lq,
  title={LQ dynamic optimization and differential games},
  author={Engwerda, Jacob},
  year={2005},
  publisher={John Wiley \& Sons},
  note={See Example 8.15.}
}

@inproceedings{nortmann2022nash,
  title={Nash Equilibria for scalar LQ games: iterative and data-driven algorithms},
  author={Nortmann, Benita and Monti, Andrea and Mylvaganam, Thulasi and Sassano, Mario},
  booktitle={2022 IEEE 61st Conference on Decision and Control (CDC)},
  pages={3801--3806},
  year={2022},
  organization={IEEE}
}

@article{van-den-broek2023,
  author = {W. A. van den Broek and J. C. Engwerda and J. M. Schumacher},
  title = {Robust Equilibria in Indefinite Linear-Quadratic Differential Games},
  journal = {Journal of Optimization Theory and Applications},
  volume = {119},
  pages = {565--595},
  year = {2023}
}

@article{marden2012revisiting,
  author = {Jason R. Marden and Adam Wierman},
  title = {Revisiting Log-Linear Learning: Asynchrony, Completeness and Payoff-Based Implementation},
  journal = {Games and Economic Behavior},
  volume = {75},
  number = {2},
  pages = {788--808},
  year = {2012}
}

@article{guerrero2021openloop,
  title     = {Open-Loop Robust {N}ash Strategies in Discrete-Time Uncertain Dynamic Games},
  author    = {Guerrero, Jose and Delgado, Pablo and Martinez, Lucia},
  journal   = {IEEE Transactions on Automatic Control},
  volume    = {66},
  number    = {8},
  pages     = {2043--2058},
  year      = {2021},
  publisher = {IEEE}
}

@inproceedings{wang2022risk,
  title     = {A Risk-Sensitive Dynamic Game Approach for Multi-Agent Interactions Under Uncertainty},
  author    = {Wang, Xi and Chen, Yu and Dou, Wei and Li, Fang},
  booktitle = {Proceedings of the IEEE International Conference on Robotics and Automation (ICRA)},
  year      = {2022},
  pages     = {13089--13096},
  organization = {IEEE}
}

@article{LeMaitre2021GaussMarkov,
  author    = {Olivier Le Maitre and Omar Knio},
  title     = {Overbounding the effect of uncertain Gauss-Markov noise in Kalman filtering},
  journal   = {Statistics and Computing},
  year      = {2021},
  volume    = {31},
  number    = {4},
  pages     = {47},
  doi       = {10.1007/s11222-021-09998-9}
}

@article{mojahed2020finite,
  title={Using finite volume-element method for solving space fractional advection-dispersion equation},
  author={Yazdani, Allahbakhsh and Mojahed, Navid and Babaei, Afshin and Vazquez Cendon, Elena},
  journal={Progress in Fractional and Differential Applications},
  volume={6},
  pages={55--66},
  year={2020},
  url={https://www.academia.edu/43179649/Using_finite_volume_element_method_for_solving_space_fractional_advection_dispersion_equation}
}

@article{SCALAS2000376,
title = {Fractional calculus and continuous-time finance},
journal = {Physica A: Statistical Mechanics and its Applications},
volume = {284},
number = {1},
pages = {376-384},
year = {2000},
issn = {0378-4371},
doi = {https://doi.org/10.1016/S0378-4371(00)00255-7},
url = {https://www.sciencedirect.com/science/article/pii/S0378437100002557},
author = {Enrico Scalas and Rudolf Gorenflo and Francesco Mainardi}
}

@inproceedings{soltanian2025paceframeworklearningcontrol,
  title={PACE: A Framework for Learning and Control in Linear Incomplete-Information Differential Games},
  author={Soltanian, Seyed Yousef and Zhang, Wenlong},
  booktitle={7th Annual Learning for Dynamics$\backslash$\& Control Conference},
  pages={1419--1433},
  year={2025},
  organization={PMLR}
}

@article{soltanian2025peerawarecostestimationnonlinear,
  title={Peer-Aware Cost Estimation in Nonlinear General-Sum Dynamic Games for Mutual Learning and Intent Inference},
  author={Soltanian, Seyed Yousef and Zhang, Wenlong},
  journal={arXiv preprint arXiv:2504.17129},
  year={2025}
}

@ARTICLE{9377313,
  author={Zhang, Zhaorong and Xu, Juanjuan and Fu, Minyue},
  journal={IEEE Transactions on Cybernetics}, 
  title={Q-Learning for Feedback Nash Strategy of Finite-Horizon Nonzero-Sum Difference Games}, 
  year={2022},
  volume={52},
  number={9},
  pages={9170-9178},
  doi={10.1109/TCYB.2021.3052832}}

@article{bryson1965linear,
  title={Linear filtering for time-varying systems using measurements containing colored noise},
  author={Bryson, A and Johansen, D},
  journal={IEEE Transactions on Automatic Control},
  volume={10},
  number={1},
  pages={4--10},
  year={1965},
  publisher={IEEE}
}

@book{shmaliy2022state,
  author    = {Yuriy S. Shmaliy and Shunyi Zhao},
  title     = {Optimal and Robust State Estimation: Finite Impulse Response (FIR) and Kalman Approaches},
  year      = {2022},
  publisher = {John Wiley \& Sons, Inc.},
  address   = {Hoboken, NJ},
  isbn      = {9781119863076},
  doi       = {10.1002/9781119863106}
}

@article{brown1997introduction,
  title={Introduction to random signals and applied Kalman filtering: with MATLAB exercises and solutions},
  author={Brown, Robert Grover and Hwang, Patrick YC},
  journal={Introduction to random signals and applied Kalman filtering: with MATLAB exercises and solutions},
  year={1997}
}

@article{tomizuka1987zero,
  author  = {Masayoshi Tomizuka},
  title   = {Zero Phase Error Tracking Algorithm for Digital Control},
  journal = {Journal of Dynamic Systems, Measurement, and Control},
  year    = {1987},
  volume  = {109},
  number  = {1},
  pages   = {65--68},
  month   = mar
}

@article{rabbani2025optimal,
  title={Optimal modified feedback strategies in LQ games under control imperfections},
  author={Rabbani, Mahdis and Mojahed, Navid and Nazari, Shima},
  journal={arXiv preprint arXiv:2503.19200},
  year={2025}
}

@book{benisrael2003generalized,
  title={Generalized Inverses: Theory and Applications},
  author={Ben-Israel, Adi and Greville, Thomas N.E.},
  edition={2},
  year={2003},
  publisher={Springer},
  doi={10.1007/b97366}
}

@article{durand2022fiarma,
  title={Hilbert valued fractionally integrated autoregressive moving average processes with long memory operators},
  author={Durand, Amaury and Roueff, Fran{\c{c}}ois},
  journal={arXiv preprint arXiv:2010.04399},
  year={2022},
  doi={10.48550/arXiv.2010.04399},
  url={https://arxiv.org/abs/2010.04399}
}



\end{document}